\documentclass[conference, letter]{IEEEtran}
\IEEEoverridecommandlockouts

\usepackage{caption}

\usepackage{comment}

\usepackage{cite}
\usepackage{amsmath,amssymb,amsfonts}
\usepackage{algorithmic}
\usepackage{graphicx}
\usepackage{textcomp}
\usepackage{xcolor}
\usepackage{framed}
\renewcommand{\theenumi}{\Alph{enumi}}
\usepackage{multirow}
\usepackage{colortbl}
\usepackage{booktabs}
\usepackage{rotating}
\usepackage{adjustbox}
\usepackage{subcaption}
\usepackage{enumitem}
\usepackage{multirow}

\def\BibTeX{{\rm B\kern-.05em{\sc i\kern-.025em b}\kern-.08em    T\kern-.1667em\lower.7ex\hbox{E}\kern-.125emX}}

\usepackage{acronym}
\usepackage{multirow}

\graphicspath{
  {./figs/}
}

\begin{document}

\title{Bug Analysis Towards Bug Resolution Time Prediction}


\author{
\IEEEauthorblockN{
Hasan Yagiz {\"O}zkan\IEEEauthorrefmark{1},
Poul E. Heegaard\IEEEauthorrefmark{2},
Wolfgang Kellerer\IEEEauthorrefmark{1},
Carmen Mas-Machuca\IEEEauthorrefmark{3},\\
}
\IEEEauthorblockA{
\IEEEauthorrefmark{1}Technical University of Munich, Chair of Communication Networks, Germany \\
\IEEEauthorblockA{\IEEEauthorrefmark{2}NTNU - Norwegian University of Science and Technology, Department of Information Security and Communication Technology,\\
}
\IEEEauthorrefmark{3} University of the Bundeswehr Munich, Chair of Communication Networks, Germany
}\\
}
\maketitle

\begin{abstract}

Bugs are inevitable in software development, and their reporting in open repositories can enhance software transparency and reliability assessment. This study aims to extract information from the issue tracking system Jira and proposes a methodology to estimate resolution time for new bugs. The methodology is applied to network project ONAP, addressing concerns of network operators and manufacturers. This research provides insights into bug resolution times and related aspects in network softwarization projects.

\end{abstract}




\section{Introduction}
\label{sec:intro}


Network softwarization entails replacing dedicated hardware for specific applications with programmable components, leading to advantages like easy maintenance, flexibility, scalability and decreased costs associated with development and operations\cite{macedo2015programmable}. Software Defined Networking (SDN) and Network Function Virtualization (NFV) are examples of these softwarization efforts. In this context, manufacturers rely on open-source solutions to provide extra functionalities while complying with the standards.
However, despite the advantages of network softwarization, operators are reluctant to use these solutions as the software is more complexity, incurring to more bugs, as well as the code reliability is uncertain as it is open to any software developer(s) differing in terms of programming skills, experience, trustiness, etc.   

In order to increase the transparency of software reliability, any bugs which are encountered during the operation of a software, are reported in open repositories. These repositories allow knowing the reported bugs, their status, etc., which are useful to evaluate the software reliability. Each software project has defined a bug resolution process, which should be followed by any software developer (i.e., debugger) of that project. 

The goal of this study is first of all, give an overview of the information that can be extracted from these repositories. Some information can be directly obtained from the gathered data, but some other need some processing. Secondly, we propose a methodology that can be used to estimate the resolution time of new bugs. The proposed methodology has been applied to several network related projects such as ONOS, ONAP, etc. We believe that network operators and manufacturers using an open source project like those, are concerned about the bug resolution times and other aspects addressed in this paper.  



The contributions of this paper are:
\begin{itemize}
    \item Collection strategy of reported bugs and the required filtering.
    \item Basic data analysis such as identification of the most common paths in the workflow, the transition distribution between different states, bug resolution time distribution, etc.
    \item Advance data analysis e.g., impact of aspects such as priority, reporter to the debugging time. 
    \item Investigating state probabilities of the bugs in time.
    \item Comparison of the bug resolution time prediction with different strategies
    \item Predicting the exact bug resolution time for different bugs with neural network.
\end{itemize}
The paper is organized as follows: Section II gives an overview of the state of the art on bug data analysis and bug resolution time models. Section III introduces the bug repositories and the data that can be extracted from them. Section IV presents the proposed bug collection and the data analysis used in this work. Section V proposes a machine learning model to estimate the bug resolution time of new bugs. Finally, Section VI concludes the paper.

\section{Related Work}
\label{sec:background}

This section will discuss previous works on bug data analysis and bug resolution processes. 

Issue tracking systems are mined, and researchers of \cite{ortu2015jira} use this process to offer the dataset of all the issues from different projects. Additionally, the authors of \cite{vieira2019reports} conduct various analyses of the collected data, which comes from 55 different Apache projects. These 55 projects are classified into 9 categories, such as big data, security, and libraries. The authors examine the percentage of fields that change the reporting of the bugs. They define bug resolution duration time from the creation and closure of a bug as a bug-fix effort and compare the bug resolution times across different project categories. Additionally, they investigate when the priority and assignee are defined and analyze the number of bugs in which these values change after the initial bug report. Furthermore, the study delves into the changes in bug status, analyzing the percentage of bugs transitioning between states. This dataset is used in this work. As the primary contribution of this work is the provided dataset, it does not delve into a comprehensive analysis or propose a methodology to estimate bug resolution duration. In addition to the self-collected data, the bug reports from \cite{vieira2019reports} are also examined in this study.

The authors investigate the impact of bug attributes such as reporter and priority on the time it takes to resolve them in the study \cite{zhang2013predicting}. The used data in this work is not publicly accessible. They employed the K-nearest neighbour algorithm to predict the resolution time of bugs. Based on these attributes from the issue-tracking system, they predicted whether a bug would be resolved quickly or slowly. The authors of \cite{akbarinasaji2018predicting} replicated the study presented in \cite{zhang2013predicting}, but they utilized open-source data sourced from Bugzilla Firefox. Both studies concluded that the identity of the reporter or assignee is more crucial than the bug's priority. However, a weakness of these predictions is that they are binary (quick or slow) rather than numerical. 

Furthermore, their proposal incorporates a Markov-based approach to estimate the number of bugs that can be resolved within a specific timeframe. They also employ a Monte Carlo-based method to predict the overall time required to fix a given quantity of bugs. The time is divided into months, and they estimate the percentage of bugs that will be resolved within a particular time period. It is important to note that the Markov-based model they used does not focus on examining the solution process of individual bugs as separate entities but rather on determining the percentage of bugs in different states after a certain period of time, measured in months.

\section{Data repositories}
\label{sec:data}

Issue tracking systems are software tools used to manage software projects' development, maintenance and debugging processes \cite{ortu2015jira}. These systems use issues which can describe bugs, feature requests, milestones and any other tasks, depending on the requirements of the considered project. Moreover, issue tracking systems also serve as a repository for the history of reported and addressed issues. Thus, they offer an excellent opportunity to understand the software project's development and debugging process. Most open-source software use publicly available issue-tracking systems because it facilitates collaboration among software developers and debuggers by giving access to the status, process, and resolution of the reported issues.

\begin{figure}
     \centering
     \includegraphics[width=\columnwidth]{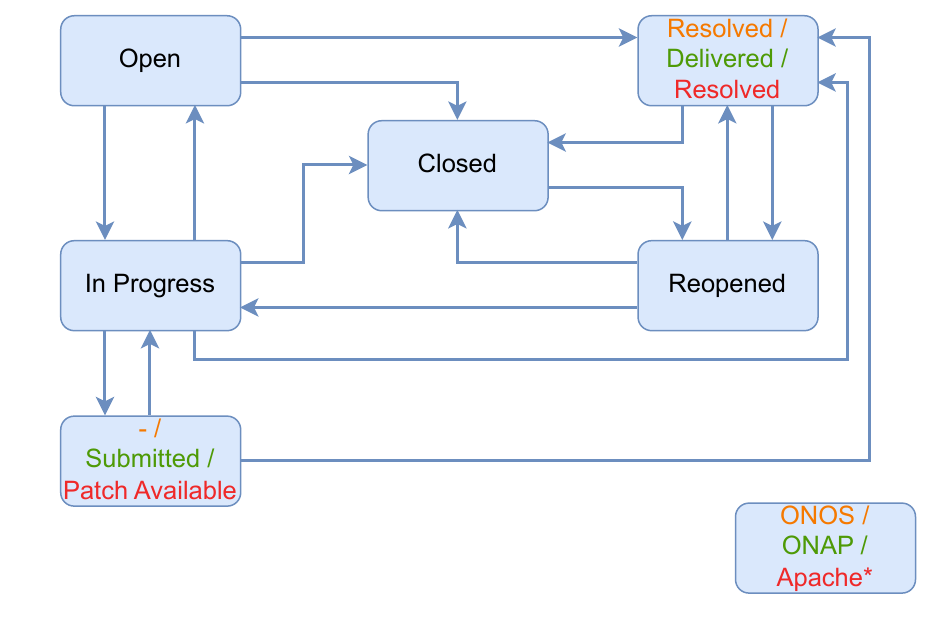}
     \caption{Comparison of ONOS, ONAP, and Apache Workflows. The ONOS workflow is fully compliant with the standard Atlassian model. The Apache* workflow applies to the considered projects.}\label{fig:transitions_onap}
\end{figure}

Jira is one of the most popular issue-tracking systems used by many different open-source projects\cite{fisher2013utilizing}. As an issue tracking system, the issue types in Jira extend beyond 'Bug,' including 'Task,' 'Story,' and more. Every issue is kept in the Jira repository of the respective project. Therefore, Jira repositories become a valuable information source for the project's development history. Jira defines a workflow, which comprises of a sequence of stages that issues may undergo, from their creation (referred as "Open") and ending with their resolution (referred as "Closed"). The workflow is depicted in Fig.~\ref{fig:transitions_onap} in blue color. 
Although Jira provides this initial workflow, it can be customized to meet the specific requirements of a project. 

This study considers two open-source projects as examples: Open Network Operation System (ONOS) and Open Network Automation Platform (ONAP). ONOS, developed by the Open Network Foundation, is an operating system designed for software-defined networking (SDN) and follows a standard workflow. ONAP is a Network Function Virtualization (NFV) Management and Orchestration (MANO) framework developed by the Linux Network Foundation~\cite{yilma2020benchmarking}. ONAP involves collaboration among various working groups. ONAP community modified the standard workflow of Atlassian to fit their bug-solving strategy by adding a state referred as "Submitted". Standard Atlassian workflow and modifications of ONAP is shown in Fig.~\ref{fig:transitions_onap}.

Furthermore, the dataset provided by \cite{vieira2019reports} is also analized. This dataset includes bug reports from 55 Apache projects. These Apache projects also use a modified version of the standard Atlassian workflow, which can be seen in Fig.~\ref{fig:transitions_onap}. This modified workflow has an additional state called 'Patch Available', similar to the 'Submitted' state of ONAP.

\subsection{Bug related data}
\label{sec:bugdata}

In this work, we are interested in the bug resolution time and hence, all issues other than 'Bug' are filtered out. Jira provides a collection of reported bugs. This collection contains details about the bug and its resolution process. Certain information is provided when a bug is reported, while others change throughout the resolution process. The following information is utilized in our analysis:

\begin{itemize}
    \item $Priority$ indicates the importance of the bug. The priority levels vary between projects. Hence, in this work, priorities $1$ and $2$ refer to the highest and second highest priority of any project.
    \item $Reporter$ is the person reporting the bug.  
    \item $Assignee$ is the person who is responsible for fixing the bug.
    \item $Creation$ $Time$ is assigned to the date and time when the bug was reported.
    \item $State$ corresponds to one of the different states of the workflow. The initial state when reporting a new bug is "Open". The state is updated during the resolution process, according to the state in the workflow.
    \item $Resolution$ $Status$ reflects the progression of the resolution process, initially set as "Unresolved" and subsequently updated based on outcomes such as "Fixed" or "Cannot Reproduce" upon completion of bug resolution.
    \item $Update$ $Time$ is the time of the last edit
    \item $State$ $Transition$ includes the previous and next states and the time of the transition.
\end{itemize}

Any modification to bug-related information in Jira (such as status, states, or additional comments) is collected, along with the timestamp and the person responsible for the update. This allows us to determine the number of people involved in the resolution process or the calculation of the time between commits and state transitions. The collected state transitions are employed to calculate the duration in each state until the bug is resolved.




\subsection{Bug resolution process} 

As bugs in Jira are a specific type of issue, they also follow the workflow for the issues. While other transitions from Fig.~\ref{fig:transitions_onap} are also utilized, a primary path within the workflow is typically followed for the bug resolution process. A bug begins in the 'Open' state when it is initially reported. Once a developer starts working on the bug resolution, the state is changed to 'In Progress'. The state of the bug changes to the 'Resolved' state, once the bug is resolved, where it awaits approval by another team member. After approval, the bug is closed by moving to the 'Closed' state. However, a bug may reappear so that it can be reopened and returned to the 'Reopen' state.

\section{Bug Collection and Analysis}
\label{sec:data_analysis}

A developed Python script has been deployed for the proposed bug collection and analysis. It is important to mention that the bugs for ONAP and ONOS were collected directly from Jira. However, the bugs for Apache were collected from the dataset at \cite{vieira2019reports}. The data analysis included in this work focuses on the two top-priority bugs (i.e., priority 1 and 2).

\subsection{Bug collection and filtering}

The collected data in this study includes the reported bugs from the start of each project to the collection date (in this case, 05.04.2023). The collected data encompasses all bug related details (refer to Section~\ref{sec:bugdata}) as well as all related information during the bug resolution process such as transition states and times, authors involved in each change, etc.

In this study, we investigate the bug resolution process. However, the dataset does not only include the data relevant to our study, but it also includes exceptional/unwilling data, which must be filtered.
Let us introduce the proposed data filtering on the dataset:
\begin{itemize}
    \item {\sl Not Resolved Bugs}: 
   Not all reported bugs are necessarily resolved; some of them may be closed for alternative reasons. The resolution status is determined as a result of the debugging process. In the case of ONAP and ONOS bugs, the resolution status begins as 'Unresolved' and subsequently changes to one of the multiple resolution statuses, which are listed in Table~\ref{tab:resolution}. The table also presents the respective percentages of each status. As this study targets the resolution time of bugs, only the bugs with the status 'Done' and 'Fixed' are considered, which account for more than 80\% of the priority 1 and 2 bugs for both ONAP and ONOS projects. Regarding the Apache dataset, no filtering is necessary as the considered dataset contains only bugs with a resolution status of 'Fixed'.
    \item {\sl Transient State}: As depicted in Fig.~\ref{fig:transitions_onap}, the resolution process consists of a set of transitions between different states, starting at "Open" and ending at "Closed". For the bug resolution time study, it is relevant to know the time at each state before any transition. However, in certain cases, bugs may spend only a few minutes or even seconds in a particular state. This short duration might not provide accurate information about the actual effort expended in that state. Transient states of this nature can arise due to automated tasks or the practice of updating Jira after completing multiple states. Consequently, states with a duration shorter than a few minutes 
    are removed as a state, and the time is added to the original state. 
    \item {\sl Not Defined States}: Observations of the dataset revealed that a few bugs have states that are not defined within the workflow of the project. The number of such states is significantly lower than those within the Jira workflow. Since the states are not in the workflow, we filter them out. The duration spent in these states is accumulated with the previous state, ensuring that the resolution duration remains unaffected.
    \item {\sl Loop to the same State}: There is a possibility that a bug has a transition from a state to the same state (referred as loop). Some of these transitions are defined as such in Jira, while others may arise due to the removal of transient states or undefined states. In the case of a loop, no additional transition is considered, and the time spent in these states is merged. 
    \item {\sl Transitions from "Close" State}: The goal of finding the bug resolution time implies that bugs eventually reach the "Closed" state. While some bugs may occasionally be reopened after a certain period, it is often observed that reopening occurs after a substantial amount of time, which significantly inflates the resolution time. To address this issue, the "Closed" state is treated as an absorbing state, meaning that once a bug reaches this state, any subsequent transitions from the "Closed" state are excluded.
\end{itemize}

\begin{table}[htb]
    \small
    \centering
    \caption{Resolution statuses and their percentage for different priority levels for ONOS and ONAP. The table excludes the Apache project dataset, which only includes bugs with status 'Fixed'.}

    \begin{tabular}[width=\columnwidth]{lcccc}
    \toprule
        Resolution& ONAP & ONAP & ONOS & ONOS\\
        Status  & Pri. 1 & Pri. 2 & Pri. 1 & Pri. 2\\
         \midrule
         Done & $\textbf{89.3\%}$ & $\textbf{87.9\%}$ & $\textbf{69\%}$ & $\textbf{60.7\%}$\\
         Fixed & $-$ & $-$  & $\textbf{27.4\%}$ & $\textbf{23\%}$\\
         Not a Bug & $3.7\%$ & $3.1\%$ & $-$ & $-$\\
         Duplicate & $2.5\%$ & $2.5\%$ & $-$ & $2.6\%$\\
         Unresolved & $2.1\%$ & $1.5\%$ & $3.5\%$ & $8.9\%$\\
         Won't Do & $1.1\%$ & $3.3\%$ & $-$ & $0.5\%$\\
         Cannot Reproduce & $1\%$ & $1.4\%$ & $-$ & $2.6\%$\\
         Not Done & $0.3\%$ & $0.3\%$ & $-$ & $-$\\
         Recommended & $0.1\%$ & $-$ & $-$ & $-$\\
         Won't Fix & $-$ & $-$ & $-$ & $1\%$\\
         Declined & $-$ & $-$ & $-$ & $0.5\%$\\
         \bottomrule
    \end{tabular}
    \label{tab:resolution}
\end{table}

\subsection{Bug resolution time distribution and common flows in transition workflow}

The resolution time is defined as the time elapsed from the reporting of a bug (i.e., in the "Open" state) to the time when it reaches the "Closed" state. The resolution time is the sum of the time spent on various states, but it does not provide information about the duration spent on each individual state.

In this section, we introduce the resolution time analysis for the particular case of the priority 1 bugs of the ONAP project.  These priority 1 bugs amount to a total of 906 in the project. 
The most common resolution paths for all the ONAP priority 1 bugs, can be found, as summarized in Table~\ref{tab:cnt}.
It can be observed in this case, that almost $91\%$ of the bugs follow the seven main paths. Furthermore, $45\%$ of the bugs follow the path \textit{ Open-In Progress-Closed} or \textit{ Open-Closed}.

\begin{table}[htb]
    \small
    \centering
    \caption{Most common debugging flows for ONAP priority 1 bug. Update values.}
    \label{tab:cnt}

    \begin{tabular}[width=0.9\columnwidth]{lll}
    \toprule
        Percentage& Flow \\
        \midrule
        25\% &  Open-In Progress-Closed \\
        20\% &  Open-Closed \\
        11\% &  Open-Delivered-Closed \\
        9\% &  Open-In Progress-Submitted-Delivered-Closed \\
        8\% &  Open-In Progress-Delivered-Closed \\
        7\% &  Open-Submitted-Closed \\
        7\% &  Open-In Progress-Submitted-Closed \\
         \bottomrule
    \end{tabular}
\end{table}


The time spent at each state before any possible transition can also be found. These time distributions are crucial for bug resolution time estimation from a specific state to the 'Closed' state. For this purpose, Table~\ref{tab:states} provides the mean and median time spent in a state before a particular transition to a different state, as well as the number of such transitions. 
Upon close examination of the transitions from the 'Open' state, it becomes apparent that the duration is longer when a bug transitions directly to the 'Closed' or 'Delivered' states compared to transitioning to the 'In Progress' or 'Submitted' states. One possible explanation for this phenomenon is that bugs typically progress through these states, even though they may not be explicitly recorded in Jira. 
The significant difference between the mean and median time for most of the states is primarily attributed to the fact that most of the bugs have a relatively short duration in each state. However, some bugs are more challenging to handle or the assignee starts later to  work on them, which causes outliers and significantly increases the mean values.

\begin{table}[htb]
    \small
    \centering
    \caption{ONAP state transition duration for priority 1 bugs}
    \label{tab:states}

    \begin{tabular}[width=\columnwidth]{lllll}
    \toprule
        From& To& Time& Time& Number \\
        & & (Mean)& (Median)&  \\
        \midrule
        Open & Closed & 472.91 & 73.2 & 143\\
        Open & In Progress & 159.01 & 26.42 & 545\\
        Open & Delivered & 361.15 & 70.47 & 85\\
        Delivered & Closed & 73.98 & 4.96 & 618\\
        Delivered & Reopened & 77.86 & 45.2 & 48\\
        Reopened & Delivered & 64.99 & 30.85 & 8\\
        Reopened & In Progress & 27.82 & 15.16 & 24\\
        Reopened & Closed & 30.65 & 15.04 & 6\\
        In Progress & Submitted & 111.48 & 25.8 & 311\\
        In Progress & Open & 342.76 & 93.39 & 21\\
        In Progress & Closed & 261.46 & 21.37 & 143\\
        In Progress & Delivered & 37.31 & 22.56 & 31\\
        Submitted & Delivered & 80.93 & 21.05 & 278\\
        Submitted & In Progress & 170.23 & 27.92 & 26\\

         \bottomrule
    \end{tabular}
\end{table}

\subsection{Impact of different bug aspects}
\label{subsec:aspects}

This section investigates the impact of different properties of the bugs on the bug resolution time. Properties like priority level, assignee and reporter of a particular bug may impact its resolution duration. The Apache dataset encompasses various projects, but we have excluded this data to eliminate variations caused by project differences. The number of bugs in ONOS should be higher to ensure a meaningful comparison. Thus, ONAP is the preferred option for investigating the assignee's and reporter's influence. Similar to the previous case, the bugs are separated based on priority.

Let us first analyse the impact of the reporter on the bug resolution time. The reporter of the bug may impact the time given his/her experience by providing an accurate and clear bug description as well as listing all the essential information to reproduce the bug accurately.

\begin{figure}
     \centering
     \includegraphics[width=\columnwidth]{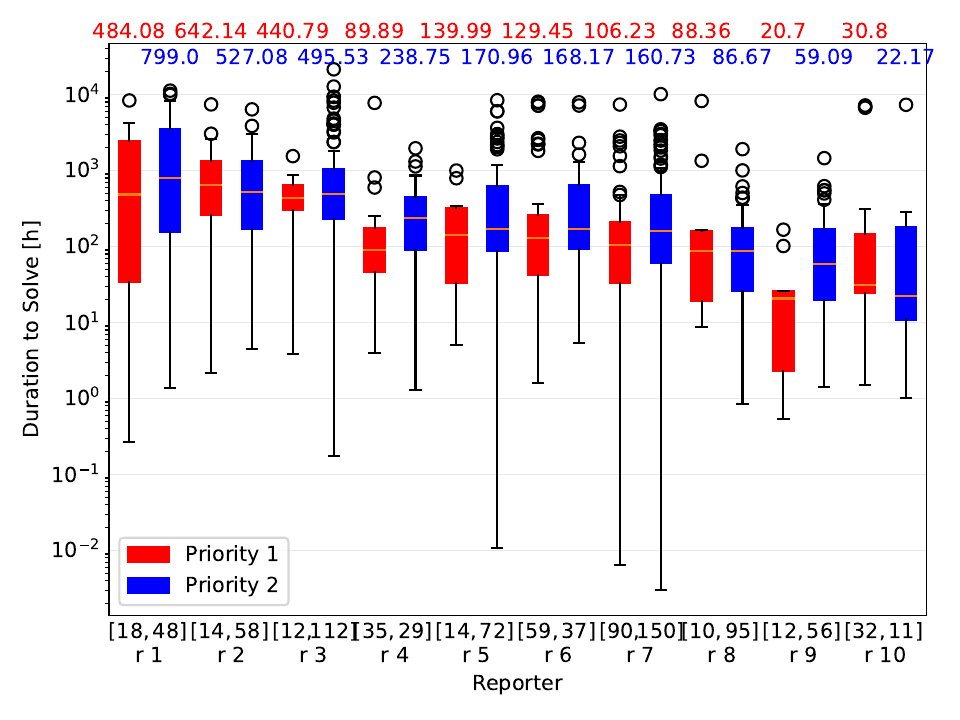}
     \caption{The role of the reporter in resolving duration of bugs in ONAP: Insights from the top ten reporters ordered by the median value of priority 2 bugs. The numbers in the x-axis represent the corresponding bug count and the median values are given on top.}\label{fig:reporter}
\end{figure}

Fig.~\ref{fig:reporter} depicts the impact of the top 10 reporters who reported the highest number of bugs, on the bug resolution time for priorities 1 and 2. The ordering of reporters is determined by the median value of the priority 2 bugs, taking into consideration that some reporters have a limited number of priority 1 bugs, with half of them reporting fewer than 20 bugs. The total number of reported bugs is shown under the x-axis of the graph, whereas the median resolution time for each case is given at the top of the graph. The decision to use the median value instead of the mean value is made in order to prioritize typical values over outliers, as the mean value can be heavily influenced by the outliers. Firstly, it is noteworthy that the resolution time for priority 1 bugs is generally shorter than that of priority 2 bugs for most reporters. Secondly, the impact of reporters on the bug resolution process is significant. In comparison to the reporter with the slowest bug resolution times, the one with the fastest experiences a median value that is 15 times quicker for priority 1 bugs and 35 times quicker for priority 2 bugs.

\begin{figure}
     \centering
     \includegraphics[width=\columnwidth]{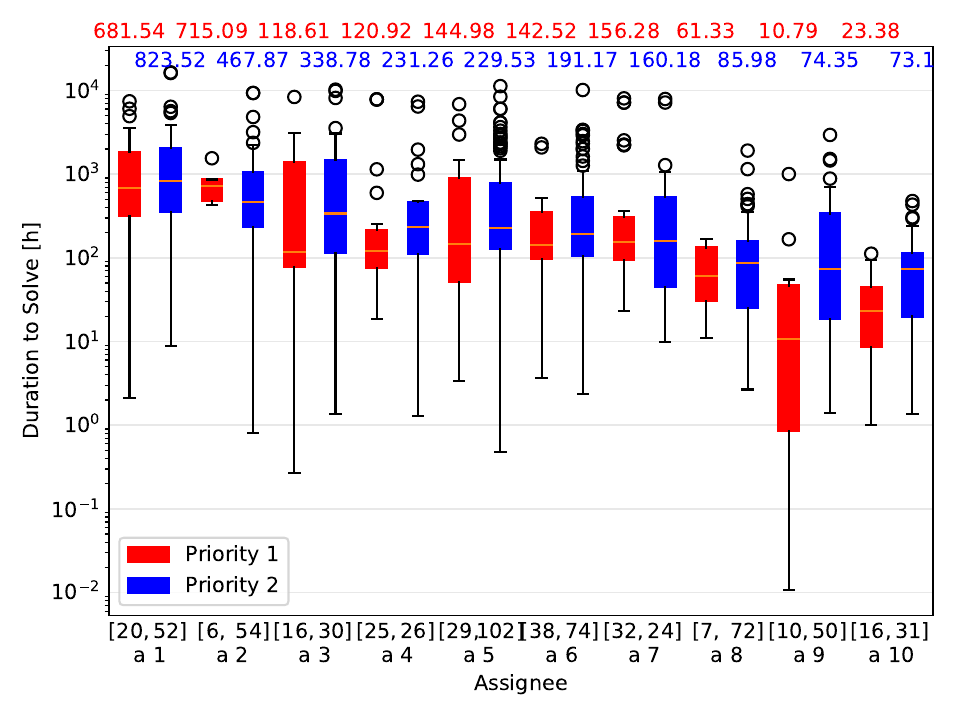}
     \caption{The role of assignees in resolving duration of bugs in ONAP: Insights from the top ten assignees ordered by the median value of priority 2 bugs. The numbers in the x-axis represent the corresponding bug count, and the mean values are given on top.}\label{fig:assignee}
\end{figure}

Another important factor is the assignee of a bug, who is responsible for resolving it. Assignees possess different expertise, availability, and familiarity with the project. The impact of various assignees is presented in Fig.~\ref{fig:assignee}. Similar to reporters, the top 10 assignees, assigned to the most bugs, are selected and ordered based on the median values of priority 2 bugs. It is evident that the assignee significantly affects the bug resolution time. Moreover, the median values on top of the figure indicate that nine out of ten authors tend to resolve higher priority bugs faster.



\begin{figure}
     \centering
     \includegraphics[width=\columnwidth]{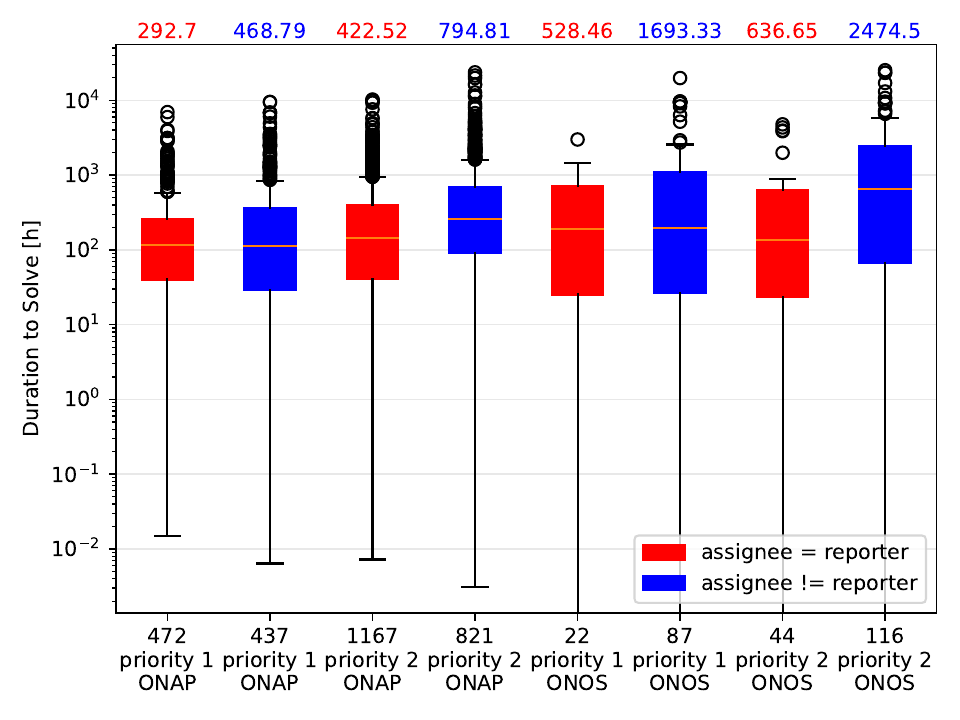}
     \caption{Impact of assigning a reporter to the bug resolution process. The numbers in the x-axis represent the corresponding bug count, and the median values are given on top.}\label{fig:assignee_reporter}
\end{figure}

The previous observations in this section indicate that the reporter and assignee play an important role in the bug resolution process. However, the reporter and assignee can be the same person, which may occur when the reporter wants to take responsibility for its resolution, e.g. when the reporter is familiar with that part of the code. In this case, it can be assumed that reproducing the bug can be easier, and there is no time required to communicate the assignee with the reporter to comprehend the bug. Thus, we further explore the impact of assigning the reporter to him/herself than to a different assignee. Fig.~\ref{fig:assignee_reporter} shows that assigning the reporter to the issue has a significant impact on bug solving process for both priority 1 and 2. 

\subsection{Bug Distribution across States over Time}
\label{subsec:state_dist}

In this section, the distribution of bug states over time is presented. It presents the probabilities of being in specific states at given times for the ONAP, ONOS, and Apache projects.

\begin{figure*}[htb!]%
\centering
\begin{subfigure}[t]{0.62\columnwidth}
\centering
\includegraphics[width=\columnwidth]{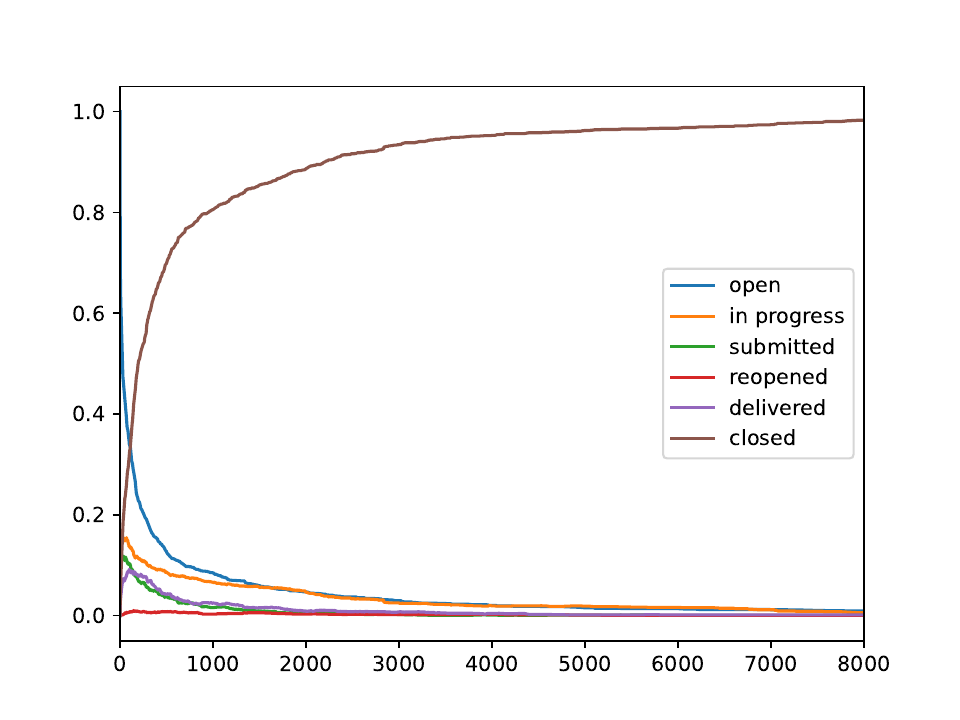}
\caption{ONAP Priority 2}%
\end{subfigure}
~ 
\begin{subfigure}[t]{0.62\columnwidth}
\centering
\includegraphics[width=\columnwidth]{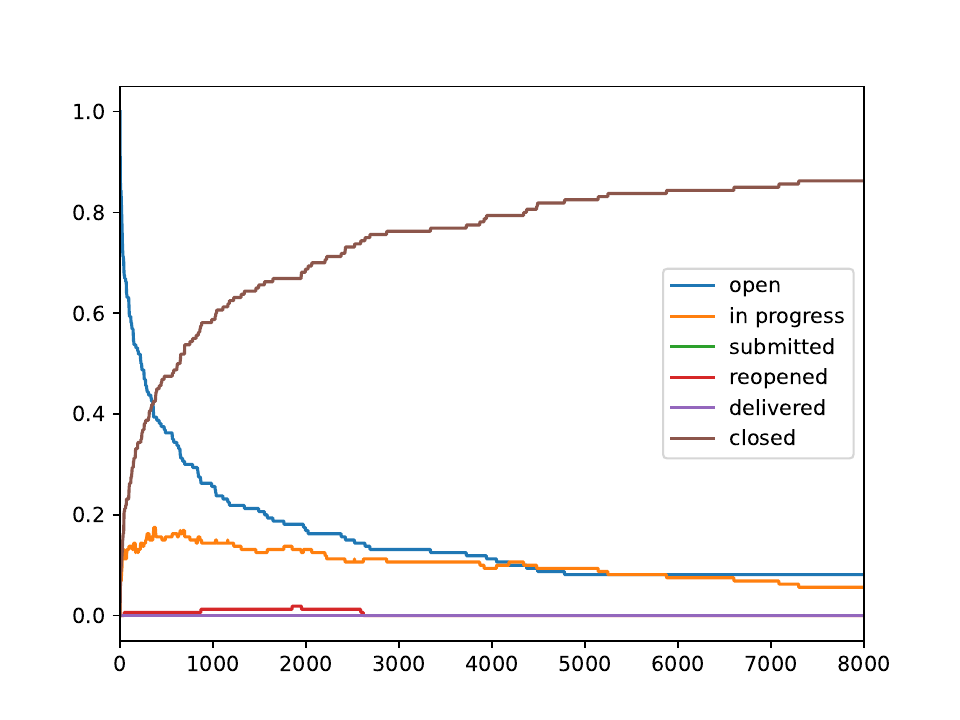}
\caption{ONOS Priority 2}%
\end{subfigure}
~ 
\begin{subfigure}[t]{0.62\columnwidth}
\centering
\includegraphics[width=\columnwidth]{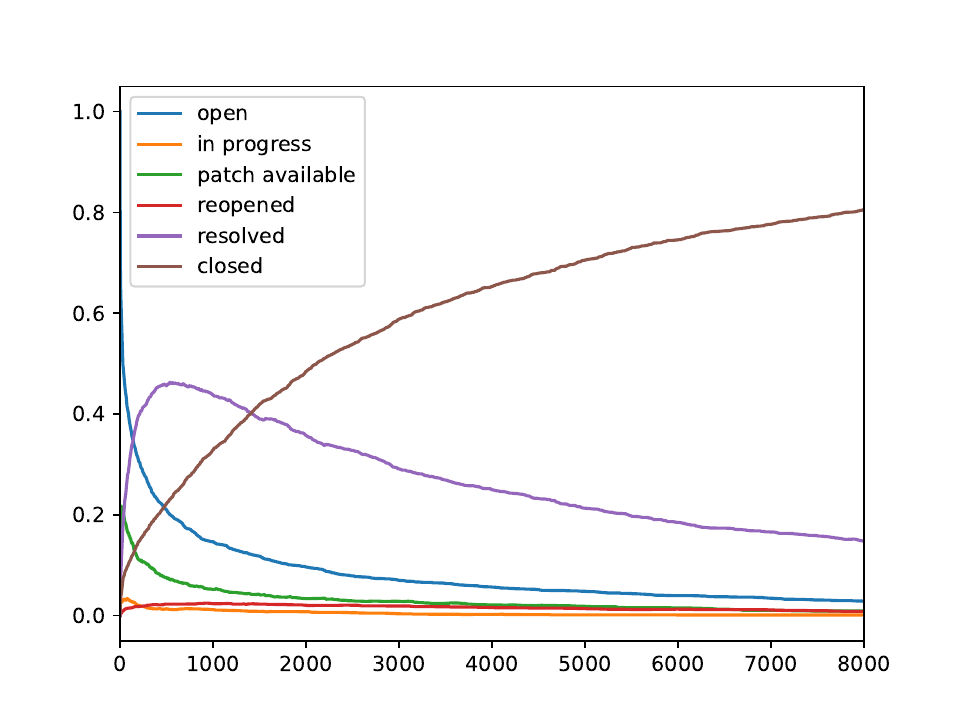}
\caption{Apache Priority 2}%
\end{subfigure}
\caption{Distribution of the bug states over time.}
\label{fig:time_state_dist}%
\end{figure*}

The time when a bug is reported (i.e., the bug is at 'Open' state), is considered to be the time 0. Hence, at this time, the initial percentage of bugs in the 'Open' state is 100\% (1.0) and zero at the other states. As time progresses, the bugs change to different states, leading to different distributions across the various states. The percentage of bugs in different states over time is provided for ONAP, ONOS and Apache projects in Fig.~\ref{fig:time_state_dist} for priorities 1 and 2. In all cases, it can be easily observed that since the 'Closed' state is absorbing, there is an always increasing percentage of bugs in that state. 

The analysis of Fig.~\ref{fig:time_state_dist} reveals a notable distinction in bug resolution times among different projects. Specifically, in ONAP, bugs consistently tend to reach the 'Closed' state within approximately $8,000$ hours, independently of the priority. Conversely, such a trend is not observed for bugs in ONOS and Apache projects, as after this time, still 10-20\% of the bugs have not reached the "Closed" state. This divergence stems from the approach adopted by Apache projects, where the emphasis is not solely on expeditiously closing bugs, but rather on considering them solved once they reach the 'Resolved' state. Additionally, in the ONOS project, some bugs remained in the 'Delivered' state for an extended period before eventually being closed.

Examining the percentage of bugs that achieve 'Closed' status, it can be observed that for priority 1 bugs, 50\% is reached after around $1,200$ hours, while for priority 2 bugs, it takes approximately $2,000$ hours. Comparatively, in ONAP and ONOS projects, more than half of the bugs are closed within $300$ hours for priority 1 bugs and within $500$ hours for priority 2 bugs. Analyzing the graph depicting bugs in the 'Closed' state, it becomes apparent that the curve for ONAP bugs exhibits a steeper ascent compared to that of ONOS and Apache projects. Additionally, it is noticeable that the number of bugs in the 'Open' state decreases more rapidly in the ONAP project. This observation suggests that the debugging teams in ONAP work on the bugs at an earlier stage in comparison to the other projects. 

However, it is important to note that making a direct comparison is challenging due to variations in the strategies employed by different working groups when transitioning between states in their respective workflows. Some projects may strictly update the bug's state as soon as it is reached in the actual handling process, while others may implement such changes after a certain amount of time has passed. This discrepancy becomes visible in the utilization of the 'Delivered' state in Fig.~\ref{fig:time_state_dist}. In Apache projects, bugs in the 'Delivered' state can be considered solved, whereas the same claim may not hold true for ONAP and ONOS projects.

\section{Predicting Resolution Time With Neural Network}
\label{sec:model}

Even after doing filterings, which are explained in Section~\ref{sec:data_analysis}, there are still bugs that can be filtered. One of the main reason to filter is the long living bugs, which are discussed in Subsection~\ref{subsec:state_dist}. Possible reasons of the short and long living bugs and different possibilities to filtering those bugs are discussed in~\cite{abdelmoez2013improving, lamkanfi2012filtering}. It has been explained that some bugs are forgotten, wrongly labeled or reported after its resolution. Therefore removing short living bugs or long living bugs might improve prediction quality. In results we compared different filtering strategies for the bugs. In~\cite{abdelmoez2013improving}, authors discuss removing long living bugs with the help of inner fence and outer fence. We also used the same strategy if all the bugs outside of inner fence is removed it is called removing mild outliers in case of all bugs above outer fence is removed is called removing extreme outliers. As an extend to these outlier removal we also proposed to remove the bugs without any update (comment, state change etc.) for more than 30 days. We compared these filtering methods and investigate their impact on the results.

We used neural networks (NN) to predict the resolution time of the bugs. The results are compared with Naive Bayesian (NB) and k Nearest Neighbour (kNN) algorithms. These two algorithms are chosen because they are widely used for resolution time prediction \cite{zhang2013predicting, 
akbarinasaji2018predicting, abdelmoez2013improving, lamkanfi2012filtering}. Since the NB and kNN cannot be applied to get numerical results but only used for classification, the results which used for comparison takes two classes into account. These two classes divided depending on the resolution time. If the resolution time is shorter than median resolution time of the training dataset, the bug considered as 'fast' bug otherwise it is classed as 'slow' bug. 

The input of the NN is the information related to the priority, subproject, assignee, reporter and either the assignee is same person of the reporter or not. Two different NN model compared for the classification one of them is trained for binary classification and returns either the bug is belongs to 'fast' or 'slow' group and the other one predicts the exact resolution time. The size and input of those models are same, where the difference lies on the loss, activation functions and the output values. For binary predicted output values are either 0 or 1 and for exact value prediction values are time to solve as hours. Since the exact time prediction is not same as binary classification, the time prediction values should be converted to two class depending on either it is predicted lower than median time or higher than median time to resolve. Both two NN consist of 5 layers, where the first one is dedicated to normalization of the input data. 

\begin{figure}
    \centering
    \includegraphics[width=\columnwidth]{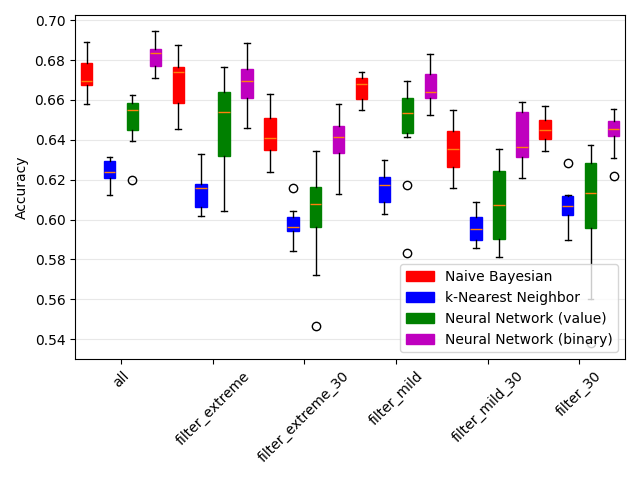}
    \caption{Comparison of the prediction of different models for ONAP. NN(values) indicates exact resolution time prediction and NN(binary) indicates the binary prediction 'slow' and 'fast' for two different NN Model}
    \label{fig:ai_comp}
\end{figure}

The ONAP project is selected for the model training, because it has more bug reports than the ONOS project and insufficient amount of data might impact the performance of the NN. In order to make prediction we divided the dataset into two group, training, which corresponds to 80\% of all dataset and test which is the rest of the dataset. In order to be able to generalize the results we cross validate our results with randomly dividing the dataset 10 times. As it can be seen from the Fig.\ref{fig:ai_comp}, even though NN does not make binary estimation of 'slow' and 'fast' its accuracy is higher than kNN estimation for any type of filtering. NN results for binary classification provides either best or second best results in any case. The filtering does not help to improve the results, one reason for that can be the reduction of the training dataset. Another possible reason is the filtered bugs mostly have similar information like assignee or subproject and classifying them as 'slow' increase the accuracy.

\begin{figure}
    \centering
    \includegraphics[width=\columnwidth]{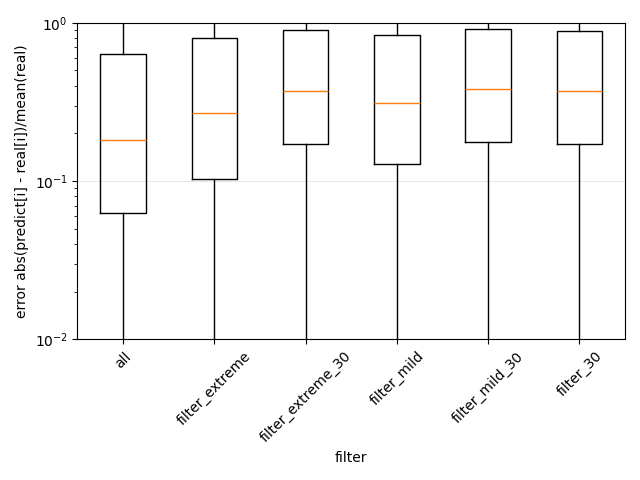}
    \caption{Comparison of the bug resolution time for .}
    \label{fig:tts}
\end{figure}

The results for exact time prediction is given in Figure~\ref{fig:tts}. It can be seen that the median normalized prediction error is 20\% for all bugs and 30\% for filtered bugs. One of the reason for this high error can be the holiday periods and weekends. The bug report day actually plays an important role on its resolution time. Even though there is an error around 20\% between predicted and real time to solve, predicted value can be more helpful than two classes.

\section{Conclusion}

In this work the importance of different aspects of the bugs on their resolution process has been shown. The time to solve a bug decrease, when the reporter is assigned to the issue. Also the impact of different resolution process or strategy to report this process has been shown. It has been seen that in Apache projects, reaching to the 'Closed' state is not forced like ONAP, instead the bugs stays in 'Resolved' state for long time. It has been shown that bug resolution time can be estimated with neural networks and the predicted value can be used by the developer teams to distribute their workforce according to the estimation. This work can be extended with using different machine learning techniques like pretraining and data augmentation to make a valid prediction for new and smaller projects.

\label{sec:conclusion}
\section*{Acknowledgment}

This work has received funding from the "DFG" under grant numbers MA 6529/4-1 and KE 1863/10-1.


\end{document}